\documentclass{aa}  

\usepackage{graphicx}
\usepackage[varg]{txfonts}
\usepackage{natbib,twoopt}
\usepackage[breaklinks=true]{hyperref}
\begin{document}

   \title{Soft X-ray absorption excess in gamma-ray burst afterglow spectra:}

   \subtitle{Absorption by turbulent ISM}

   \author{M. Tanga\inst{1} \fnmsep \thanks{E-mail: mohit@mpe.mpg.de}\and P. Schady\inst{1}\and A. Gatto\inst{2}\and J. Greiner\inst{1}\and M. G. H. Krause\inst{3,1,4}\and R. Diehl\inst{1}\and S. Savaglio\inst{1,5,6}\and S. Walch\inst{2,7}}

  \institute{Max-Planck-Institut f{\"u}r Extraterrestrische Physik, PO Box 1312, Giessenbachstr. 1, 85741, Garching, Germany             
        \and Max-Planck-Institut f{\"u}r Astrophysik, Karl-Schwarzschild-Stra{\ss}e 1, D-85748 Garching, Germany 
        \and Universit{\"a}ts-Sternwarte M{\"u}nchen, Ludwig-Maximilians-Universit{\"a}t, Scheinerstr. 1, 81679 M{\"u}nchen, Germany
        \and Excellence Cluster Universe, Technische Universit{\"a}t M{\"u}nchen, Boltzmannstrasse 2, 85748 Garching, Germany
        \and European Southern Observatory, Karl-Schwarzschild-Stra{\ss}e 2, 85748 Garching, Germany
        \and Physics Dept., University of Calabria, via P. Bucci, I-87036 Arcavacata di Rende, Italy       
        \and Physikalisches Institut, Universit{\"a}t K{\"o}ln, Z{\"u}lpicher Strasse 77, D-50937 K{\"o}ln, Germany
        }
   \date{ }

 
  \abstract{
   {}
   {}
   {}
   {}
   {}

Two-thirds of long duration gamma-ray bursts (GRBs) show soft X-ray absorption in excess of the Milky Way. The column densities of metals inferred from UV and optical spectra differ from those derived from soft X-ray spectra, at times by an order of magnitude, with the latter being higher. The origin of the soft X-ray absorption excess observed in GRB X-ray afterglow spectra remains a heavily debated issue, which has resulted in numerous investigations on the effect of hot material both internal and external to the GRB host galaxy on our X-ray afterglow observations. Nevertheless, all models proposed so far have either only been able to account for a subset of our observations (i.e. at $z>2$), or they have required fairly extreme conditions to be present within the absorbing material. In this paper, we investigate the absorption of the GRB afterglow by a collisionally ionised and turbulent interstellar medium (ISM). We find that a dense (3 cm$^{-3}$) collisionally ionised ISM could produce UV/optical and soft X-ray absorbing column densities that differ by a factor of 10, however the UV/optical and soft X-ray absorbing column densities for such sightlines and are 2-3 orders of magnitude lower in comparison to the GRB afterglow spectra. For those GRBs with a larger soft X-ray excess of up to an order of magnitude, the contribution in absorption from a turbulent ISM as considered here would ease the required conditions of additional absorbing components, such as the GRB circumburst medium and intergalactic medium.
}


   \maketitle
   
%

\section{Introduction}
Long duration gamma-ray bursts ~\citep[GRBs; duration $\gtrsim  2$s;][]{Kouveliotou1993ApJ} are thought to originate from the collapse of rapidly rotating massive stars (25-30 $M_{\odot}$)~\citep{Woosley2006ApJ} and this is supported by their observed association with supernovae (SNe) of type I b/c~\citep{Galama1998Nature, Stanek2003ApJ, Hjorth2003Nature}. With luminosities of $10^{51-52}$ erg/s over cosmological distances from $z = 0.0085$ ~\citep{Galama1998Nature} to $z > 8.0$~\citep{Tanvir2009Nature,Salvaterra2009Nature,Cucchiara2011ApJ}, and emission over most of the electromagnetic spectrum, GRBs are excellent probes of the environment of their host galaxies and any intervening absorbers ~\citep{Savaglio2003ApJ,Chen2005ApJ,Fynbo2006AA,Savagio2006NJPh,Prochaska2007ApJ,Piramonte2008AA,Fynbo2009ApJS,Savaglio2012MNRAS,Zafar2012ApJ,Japelj2015AA}. 

Since the launch of the {\it Swift} satellite mission in 2004 ~\citep{Gehrels2004ApJ} high quality X-ray spectra (0.3 - 10 keV) of GRB afterglows are now routinely obtained with the X-Ray Telescope ~\citep[XRT;][]{Burrows2004SPIE}. The rapid arc-second localisation of GRB afterglows with the XRT, which has a detection rate of $>$ 95$\%$~\citep{Evans2009MNRAS}, has enabled quick photometric and spectroscopic follow up from ground-based observatories. These follow-up observations have provided the redshift of the burst and have resulted in high signal-to-noise (S/N) observations of the absorption features originating from the circumburst medium, host galaxy interstellar medium (ISM) and intervening absorbers in UV/optical/NIR afterglow spectra. The column densities of neutral gas inferred from Lyman-$\alpha$ absorption in GRB afterglow spectra indicate that a high percentage of GRB hosts are damped Lyman-$\alpha$ systems (DLAs; $N_{HI} > 2 \times 10^{20.0}$) or sub-DLAs ($10^{19.0} < N_{HI} < 2 \times 10^{20.0}$) ~\citep{Wolfe2005ARAA,Prochaska2007ApJ,Fynbo2008ApJ,Savaglio2012AN,Sparre2014ApJ,Cucchiara2015ApJ}. Absorption features due to singly ionised species such as \ion{Zn}{ii}, \ion{Fe}{ii}, \ion{S}{ii,} and \ion{Si}{ii} from the ISM are regularly seen in the GRB UV/optical afterglow spectra. On the basis of temporarily varying lines of singly ionised species due to photo-excitation by UV afterglow photons, the absorbing gas can be located at a distance of a few 100 pc from the GRB progenitor ~\citep[e.g.][]{Vreeswijk2007AA,DElia2011MNRAS, Vreeswijk2013AA, Hartoog2013MNRAS, Kruehler2013AA}. Absorption from highly ionised species within the host galaxy, such as \ion{C}{iv}, \ion{N}{v}, \ion{O}{vi} and \ion{Si}{iv}, are more challenging to detect because they lie within the Lyman-$\alpha$ forest, and they have weaker line strengths. However, their detection has been reported in a dozen GRB afterglows at $z>2$ ~\citep{Prochaska2008ApJ, Fox2008AA}. The highly ionised species are located either within the circumburst medium of the GRB, $<$ 20 pc from the progenitor, or within the ISM at distances of  $>$ 400 pc from the GRB progenitor, and can extend into the halo of the GRB host ~\citep{Fox2008AA}. 

The soft X-ray spectra of 60$\%$ of GRBs show absorption in excess of Milky Way absorption ~\citep{Evans2009MNRAS, Campana2010MNRAS, Campana2012MNRAS, Starling2013MNRAS}. However, the nature and location of the soft X-ray absorbing material is not clear, since the narrow line absorption/emission features in soft X-ray spectra are not resolved with the {\it Swift}/XRT. Although ASCA ~\citep{Tanaka1994PASJ}, BeppoSAX ~\citep{Boella1997AAS}, Chandra ~\citep{Weisskopf2000SPIE} and XMM-Newton ~\citep{Jansen2001AA} X-ray observatories are capable of resolving narrow line features, this possibility is impeded by typically long delays before the GRB observations begin by which time the X-ray afterglow has dimmed significantly. Previously claimed X-ray absorption and emission features in pre-Swift GRB afterglow spectra ~\citep{Piro1999ApJ,Yoshida1999AAS,Antonelli2000ApJ,Piro2000Science,Yoshida2001ApJ,Reeves2003AA,Butler2003ApJ,Watson2003ApJ} have been largely put in doubt ~\citep{Sako2005ApJ}.

The interstellar gas absorbs X-rays largely independent of temperature as long as this gas is not fully ionised. Therefore the cold neutral, warm ionised and hot ionised phases of the ISM are equally opaque to soft X-rays. The equivalent hydrogen column density ($N_{HX}$) inferred from absorption in soft X-ray spectra is greater than the host neutral hydrogen column density $(N_{HI})$ in a high percentage of GRBs~\citep{Watson2007ApJ,Campana2010MNRAS} with the difference at times exceeding a factor of ten. The $N_{HX}$ for GRBs with known redshift is measured by fitting the broad soft X-ray absorption with a neutral absorber, and then converting the best-fit metal column density to an equivalent neutral hydrogen column density assuming solar abundances. GRBs preferentially occur in subsolar metallicity environments ~\citep{LeFloch2003AA, Modjaz2008AJ, Savaglio2009ApJ, GrahamFruchter2013ApJ} and if this is considered, $N_{HX}$ would increase further. The soft X-ray absorption traces the total metal content along the GRB line of sight rather than the hydrogen content and thus one should compare neutral metal column densities with the total metal column densities to estimate the excess X-ray absorption. ~\citet{Schady2011AA} compared GRB host column densities of singly and highly ionised species with the soft X-ray absorbing metal column densities rather than the neutral hydrogen column densities and concluded that these species were insufficient to account for the excess soft X-ray absorption. Instead they postulated that a significant component of ultra-highly ionised gas that is transparent to UV photons could be responsible for the excess absorption.

Other investigations into this problem have concentrated on sources of absorption that are both intrinsic and external to the GRB host galaxy. ~\citet{Campana2010MNRAS, Campana2012MNRAS} and more recently~\citet{Starling2013MNRAS}, analysed a large sample of {\it Swift} GRBs with known redshift, and found that $N_{HX}$ increased with redshift, implying that the soft X-ray absorption excess has an external origin. ~\citet{Behar2011ApJ} found that the opacity of GRBs beyond $z =$ 2 remains constant, and suggested that a low redshift neutral and diffuse intergalactic medium (IGM) with metallicity of 0.2-0.4$Z_{\odot}$ could be the X-ray primary absorber for GRBs at high redshift ($z \gtrsim 2.0$). Similarly, ~\citet{Starling2013MNRAS} conclude that a warm-hot IGM (WHIM) with temperatures between $10^{5-6}$ K and metalicity $> 0.2 Z_{\odot}$ could explain the excess absorption for GRBs at $z \gtrsim 3$. However, recent results from surveys of intergalactic \ion{O}{vi} absorbers at low redshift with HST show that the median metalicity (of IGM gas) at $T \sim 10^{5.5}$ K is 0.1$Z_{\odot}$ ~\citep{Savage2014ApJS}. On the basis of cosmological simulations ~\citet{Campana2015AA} show that metals in the circumgalactic medium (T $\sim 10^{6-7}$ K; $0.2 - 0.4 Z_{\odot}$) expelled by galactic winds from dense groups of small galaxies at low redshift can account for soft X-ray absorption in GRBs at $z \gtrsim 3$. For GRBs at $z < 2$, the contribution to the soft X-ray absorption from components external to the GRB host galaxy are expected to be $N_{HX} < 10^{22}$ cm$^{-2}$. On the other hand, ~\citet{WatsonJakobsson2012ApJ} have previously argued that selection effects may produce the $N_{HX}-z$ correlation, and when low redshift, dust obscured GRBs with high $N_{HX}$ ($> 10^{22}$ cm$^{-2}$) are included in the GRB samples, the significance of this correlation is greatly reduced, thereby suggesting that the IGM plays a lesser role in causing excess soft X-ray absorption. 

Absorption by sources internal to the GRB host have been suggested owing to the fact that $N_{HX}$ is larger for bursts with higher $N_{HI}$ ~\citep{Kruehler2011AA, Kruehler2012ApJ, Campana2012MNRAS, Watson2013ApJ}. On the basis of both a positive correlation between $N_{HX}$ and line of sight host dust extinction ($A_{V}$), and also between $N_{HX}$ and $N_{HI}$,~\citet{Watson2013ApJ} attribute the soft X-ray absorption to dense \ion{H}{ii} regions of the GRB host. Since singly ionised helium produces a soft X-ray absorption profile similar to neutral hydrogen without absorbing UV photons, ~\citet{Watson2013ApJ} suggest that \ion{He}{ii} in low metallicity, dense \ion{H}{II} regions ($10^{3-4}$ cm$^{-3}$) within 5 pc of the GRB is the dominant source of the absorption. Higher column densities of \ion{He} would increase the X-ray opacity within the \ion{H}{ii} region by almost an order of magnitude. ~\citet{Krongold2013ApJ} , who studied the time-dependant photoionisation of the circumburst environment by the afterglow, found that a soft X-ray excess $\gtrsim 10$ can be explained if the progenitor is embedded in a dense molecular cloud with a density that is $\sim 10-100$ times higher than the ISM, and is partially ionised out to $5-30$ pc. In their simulations, solar metallicities within the dense absorbing material were required to produce an order of magnitude difference in X-ray and UV absorbing column densities. However, the largest problem that these models face is in explaining how such a highly ionised dense medium could remain in pressure equilibrium with the cold and dense neutral phase ISM.

Given the continual lack of a clear explanation for the soft X-ray absorption excess, in this paper we explore the effect of hot gas within the GRB host ISM may have on our X-ray afterglow observations. Since GRBs are associated with massive stars and SNe, the ISM of such galaxies is likely to be turbulent ~\citep[e.g.][]{OstrikerShetty2011ApJ, Walch2011ApJ, Avillez2012ApJ}, with large volume filling factors of hot and ultra-highly ionised gas, as has been observed in a number of GRB hosts ~\citep{Fox2008AA}. We aim to understand the imprint that such an ISM would have on the GRB afterglow spectra, with a particular focus on how the ultra-highly ionised species traced by, \ion{C}{v}, \ion{N}{vi} and \ion{O}{vii} affect the afterglow absorption.

To do this, we consider data cubes obtained from simulations of a multiphase and turbulent ISM driven by SNe with gas that is in collisional ionisation equilibrium (CIE). We pass random lines of sight through the data cubes and keep track of the absorbing column of all C, N, and O ions along each line of sight. The paper is organised as follows: In section 2 we discuss the expected turbulence in the ISM within GRB host galaxies, provide details of the ISM data cubes that we used, and describe our line-of-sight code used to determine the hydrogen and metal column densities. In section 3 we describe how we estimate the X-ray excess absorption from the measured column densities and compare our results to observations. We discuss our results in section 4 and we provide a summary of our results in section 5.

\section{Methodology}
\begin{figure*}
\centering
        \includegraphics[width=17cm]{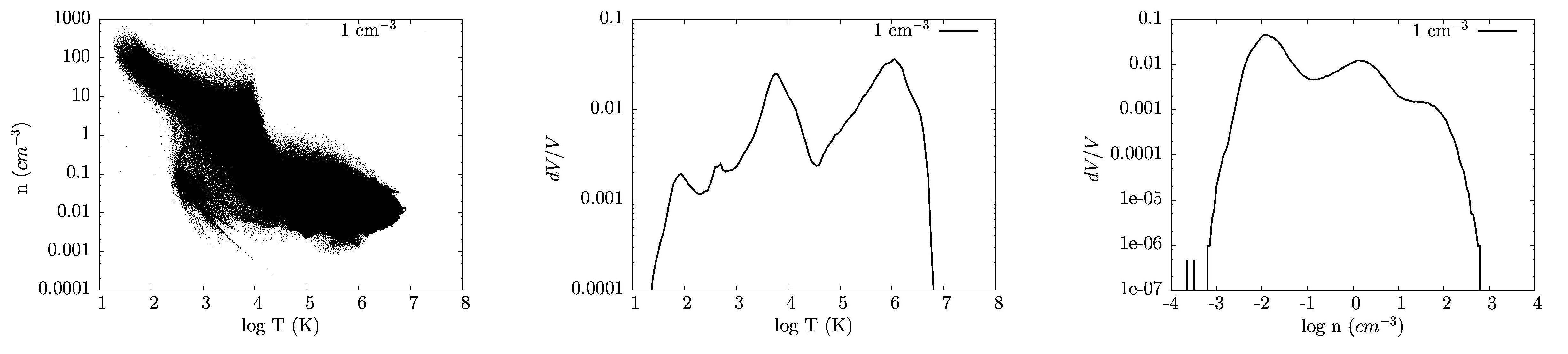}
        \caption{ISM data cube with density 1 cm$^{-3}$. Left: Particle number density as a function of temperature. Middle: Volume-weighted probability density functions (PDFs) as a function of temperature. Right: Volume-weighted PDF as a function of particle number density.}
        \label{fig1:ISM-1cc}
\end{figure*}

\begin{figure*}
\centering
        \includegraphics[width=17cm]{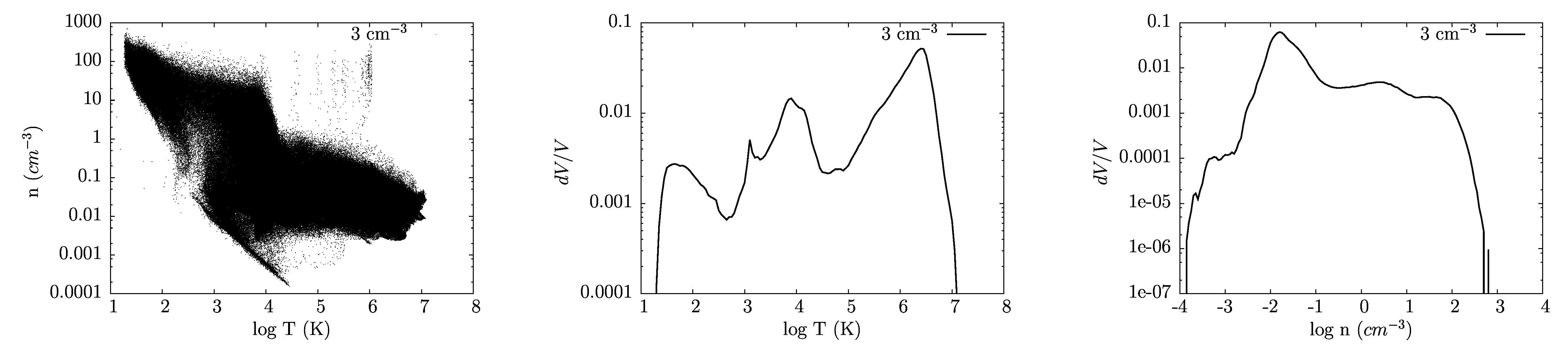}
        \caption{As in Fig.~\ref{fig1:ISM-1cc} but for the ISM data cube with density 3 cm$^{-3}$.}
        \label{fig2:ISM-3cc}
\end{figure*}

\begin{figure*}
\centering
        \includegraphics[width=17cm]{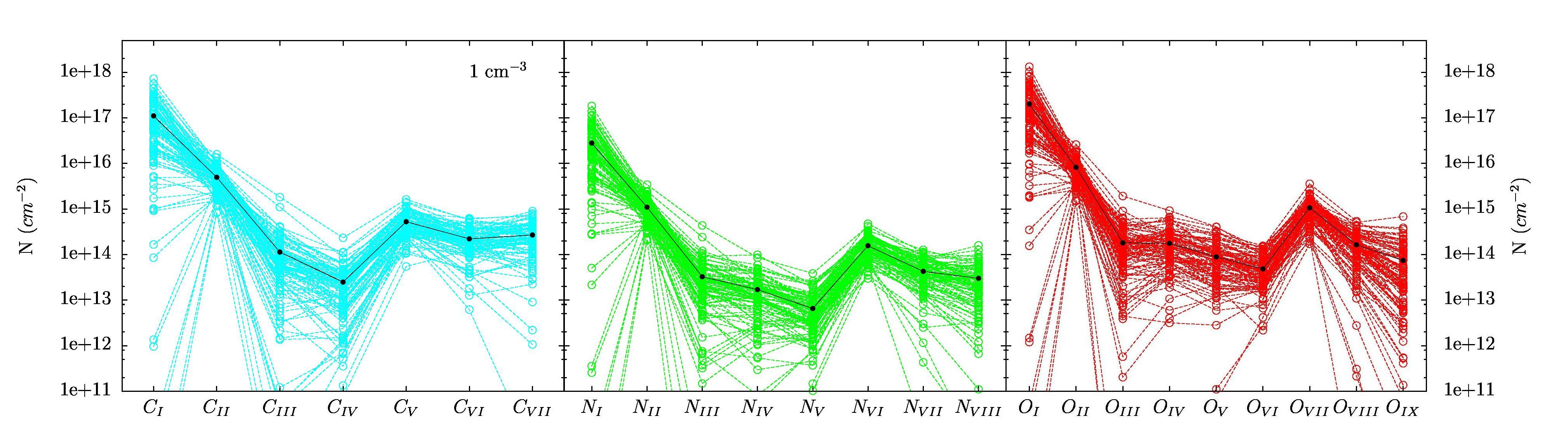}
        \caption{Column densities of C (left), N (middle) and O (right) ions for 100 sightlines passed through our ISM data cube with density 1 cm$^{-3}$ (datacube1). The black points indicate the average ion column densities for our 100 sightlines.}
        \label{fig3:CNO-ion-1cc}
\end{figure*}

\begin{figure*}
\centering
        \includegraphics[width=17cm]{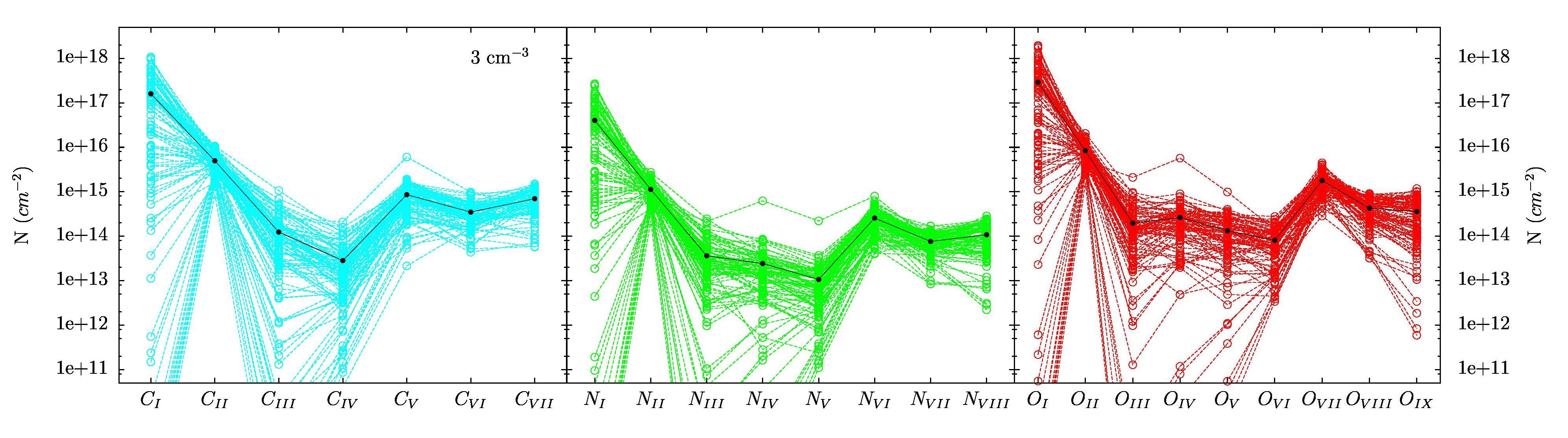}
        \caption{As in Fig.~\ref{fig3:CNO-ion-1cc} but for the ISM data cube with density 3 cm$^{-3}$.}
        \label{fig4:CNO-ion-3cc}
\end{figure*}

\subsection{Turbulent ISM: A property of low mass star-forming galaxies}
Turbulence within the ISM is driven by massive stellar feedback \citep[e.g.][]{OstrikerShetty2011ApJ, Kim2013ApJ}. GRB hosts have high SFR \citep[e.g.][]{Savaglio2009ApJ}, are associated with SN explosions and are thus likely to have a turbulent ISM.

Host galaxies of GRBs have a typical mass of $\sim$$10^{9} M_{\odot}$ and like most low mass ($< 10^{10} M_{\odot}$) star-forming galaxies (SFGs) are often seen to be irregular and clumpy ~\citep{Fruchter2006Nature}. These consist of star-forming clumps and lack a central bulge ~\citep[e.g.][]{ElmegreenElmegreen2005ApJ, FoersterSchreiber2011ApJ}. The kinematics inferred from H-$\alpha$ line emission suggest that SFGs are rotating disks or are dominated by random motion ~\citep[e.g.][]{FoersterSchreiber2006ApJ, Newman2013ApJ}. The H-$\alpha$ emission-line studies and resolved maps of  such galaxies show their ISM to be turbulent with large outflows ~\citep[e.g.][]{Newman2012ApJ}. \citet{Milvang_Jensen2012ApJ} find that GRB hosts are Lyman$-\alpha$ emitters with equivalent widths larger than bright Lyman break galaxies (LBGs). ~\citet{Fox2008AA} report all seven GRBs in their sample to be absorbed by high velocity (few 100 km/s), highly ionised gas within the host galaxy, and ~\citet{Thoene2007ApJ} report absorption by high velocity, low ionised gas clouds possibly from the halo of the host galaxy, all of which are indicative of star formation driven outflows.

Massive star winds and SN explosions inject energy into the ISM causing it to heat up and expand. These regions of hot gas form SN bubbles extending out to kpc scales ~\citep{Avillez2004AA, Krause2013AA} have a low particle density and exist in dynamic equilibrium with high density and low temperature neutral gas creating a multiphase ISM. We consider simulations of an ISM that employ these properties.

\subsection{Interstellar medium data cubes}
Details of the three-dimensional ISM data cubes that we used are given in ~\citet{Gatto2015MNRAS}. In summary, the size of the box is 256 pc$^{3}$ with a resolution of $2$ pc. The two data cubes we used for this work had an initial hydrogen density of 1 cm$^{-3}$ and 3 cm$^{-3}$, and we label them as datacube1 and datacube3 for the rest of the paper. Simulations with initial temperature of $5000$ K were driven artificially for 25 Myr prior to the onset of the SN explosions. Thereafter, turbulence in these boxes is driven by type II SN explosions, where the SNe are placed at random peaks at all times for datacube1, and at density peaks for $50\%$ of the time and at random positions for the other $50\%$ for datacube3. SN explosions at peak position correspond to SNe that explode within their natal region, while the SN explosions at random positions correspond to SN explosions from runaway stars. In the latter case, the SN explosion is more likely to occur in the ISM phase with largest volume filling factor and thus low density gas. The higher the density of the ambient medium of a SN, the more of the energy injected by the SN that is radiated away, and the lower the energy transfered to the ISM. Thus ISM with peak SN explosions causes a low fraction of hot gas within the ISM, and ~\citet{Gatto2015MNRAS} conclude that these simulations do not produce a realistic ISM. However, they further demonstrate that as the density of the ISM data cube increases to $\ge$ 3 cm$^{-3}$, purely random SN explosions cause the ISM data cubes to be completely filled with hot gas due to a lack of radiative cooling and lack of gas outflows because of the strict periodic boundary conditions employed for the simulations. We thus restrict our analysis to the data cube with density 1 cm$^{-3}$ with SNe placed at random positions, and the data cube with density 3 cm$^{-3}$ with SN explosions distributed equally between peak and random positions.

The SN rates for simulations with densities 1 cm$^{-3}$ and 3 cm$^{-3}$ are set at 3 Myr$^{-1}$ and 14 Myr$^{-1}$, respectively. The SN rate is chosen to be proportional to the star formation rate (SFR), where the latter is assumed to follow the Kennicutt-Schmidt relation ~\citep{Kennicutt1998ApJ}. Using a Chabrier initial mass function (IMF) ~\citep{Chabrier2003PASP} and star formation efficiency of $1\%$ this corresponds to one SN per $10^{2}$ $M_{\odot}$ of stellar mass formed with a predefined thermal energy of $10^{51}$ erg. The simulations with densities 1 cm$^{-3}$ and 3 cm$^{-3}$ were run for 130 Myr and 198 Myr, respectively. We use two simulations with different gas densities to see if a relation between the ISM type and  X-ray absorbing gas exists. The densities of carbon, atomic hydrogen, and molecular hydrogen are given as inputs. Solar abundance ratios are assumed for both the simulations. Although GRBs are typically found to reside within subsolar metallicity galaxies ~\citep{LeFloch2003AA,Modjaz2008AJ,Savaglio2009ApJ,GrahamFruchter2013ApJ}, for metallicities in the range $1.0 Z_{\odot} > Z > 10^{-3}Z_{\odot}$, the dynamics of the simulated ISM are a weak function of metallicity ~\citep{Walch2011ApJ}. While super-solar GRB hosts have been reported ~\citep{Levesque2010ApJ, Savaglio2012MNRAS, Kruehler2012AA, GrahamFruchter2013ApJ, Elliott2013AA}, no GRB hosts with $Z < 10^{-3}Z_{\odot}$ have been observed. The lowest metallicity thus far known for a GRB host galaxy is $Z \sim 0.01 Z_{\odot}$ (GRB050730; Starling et al. 2005, GRB090926A; Rau et al. 2010. D’Elia et al. 2010), and the average host metallicity is $\sim 0.5 Z_{\odot}$ ~\citep{Prochaska2007ApJ, Savaglio2012AN}. We therefore expect the filling factor of warm and hot gas in our data cube to be a fair representation of the multiphase ISM within GRB host galaxies.

The formation and cooling of ions (\ion{H}{I}, \ion{H}{II} and \ion{C}{II}) and molecules ($H_{2}$, CO) are calculated as the simulations using algorithms from ~\citet{Glover2010MNRAS} and ~\citet{GloverClark2012MNRAS}, and the cooling rate for highly ionised oxygen, for example \ion{O}{vi}, is incorporated in accordance with CIE ~\citep{GnatFerland2012ApJS}. Thus, our treatment of the gas under the conditions of CIE is consistent with the simulations. The gas within the injection radius is heated to $10^{6} - 10^{7}$ K when the SN explodes, and the speed of the blast expanding within the SN bubble corresponds to the speed of sound within the local region. The superbubbles then expand adiabatically and cool correspondingly. They are, however reheated time after time with subsequent explosions.

The simulations do not take the ionisation of dense regions by UV emission from massive stars into account, and only trace the collisionally ionised gas, which is only relevant above $10^{4}$ K. Thus all gas below $10^{4}$ K is approximated to be neutral. Under conditions of CIE, \ion{C}{v}, \ion{N}{vi}, and \ion{O}{vii} form at T $> 10^{5}$ K and peak at T $\sim 10^{5.5}$ K ~\citep{SutherlandDopita1993ApJS}. Large column densities of highly ionised gas can therefore only be produced in our simulations if the fraction of volume covered by these hot-diffuse regions is high. In Fig.~\ref{fig1:ISM-1cc} and Fig.~\ref{fig2:ISM-3cc} we plot the particle number density against temperature in the left panel, and the volume-weighted probability density functions (PDFs) for temperature and density, in the middle and right panels, respectively, for datacube1 and datacube3. The general widths of the temperature and density distributions is similar to that of previous studies ~\citep[e.g.][]{Avillez2004AA} .The middle and right panels of Fig.~\ref{fig1:ISM-1cc} and Fig.~\ref{fig2:ISM-3cc} show that the hot-diffuse regions within the ISM data cubes do indeed cover a high volume fraction. Fig.~\ref{fig2:ISM-3cc} also shows the presence of a higher volume fraction of hot and low density gas in datacube3 compared to datacube1, and a higher volume fraction of cooler gas which leads to the formation of higher densities of molecular hydrogen.

\subsection{Line of sight sampling}
Rays of random orientation are passed through 100 random positions within the ISM data cubes with typical path lengths $\sim$ 350 pc. We ensure that our sightlines sample large length scales (few 100 pc). In tracking the cells within the data cube that are crossed by our ray, only those cells where the ray crosses $> 1/10$ of the resolution of the cell (i.e. 0.2 pc) are considered, and we verify that no cell is selected twice. The temperature of the gas and number density of hydrogen is extracted for each selected cell, and the C, N, and O number density within each cell is calculated assuming solar abundances from ~\citet{Asplund2009ARAA}. To calculate the ion fractions for C, N, and O, we use the CIE tables from ~\citet{SutherlandDopita1993ApJS}, which are appropriate for solar metallicities and are consistent with the assumptions that went into the data cube simulations. The distance traversed by the ray within each cell is calculated by keeping track of the entry and exit point for the cell, and the total distance traversed by the ray is then the sum of the distances traversed through all selected cells. The column densities of H, C, N, and O ions along the line of sight is the summation of their respective column densities in each cell. It is important to calculate the distance travelled within each cell to measure the column densities accurately as neither the density within the data cube nor the distances travelled within each cell are uniform.

Amongst C, N, and O, the ionisation potentials for \ion{O}{i} and \ion{H}{i} are most similar (13.598 eV and 13.618 eV), and thus we consider \ion{O}{i} as a tracer of neutral gas and oxygen ion species from \ion{O}{i} to \ion{O}{viii} as tracers of the X-ray absorbing gas, which we refer to as $N_{OX}$. The absorption excess along each sightline is estimated as the ratio of the X-ray absorbing oxygen column density, $N_{OX}$, to neutral oxygen column density, $N_{OI}$.

\section{Results}
\begin{figure}
\centering
        \resizebox{\hsize}{!}{\includegraphics{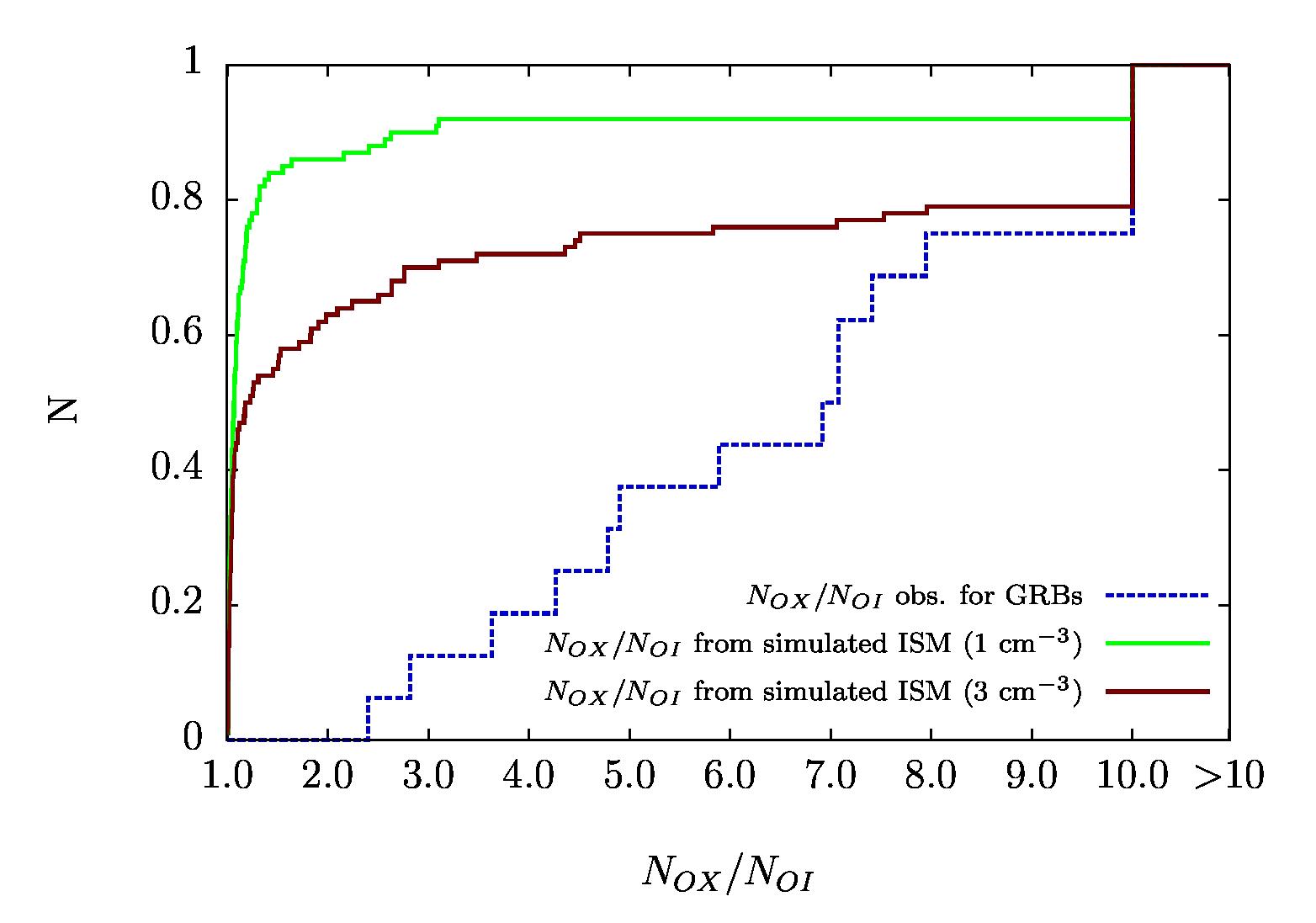}}
        \caption{Green and brown lines show the the cumulative distribution of $\frac{N_{OX}}{N_{OI}}$ for 100 sightlines crossed through our ISM data cubes with densities 1 cm$^{-3}$ and 3 cm$^{-3}$, respectively, and the blue dashed line shows the observed cumulative distribution of $\frac{N_{OX}}{N_{OI}}$ for a sample of 16 GRBs.}
      \label{fig5:O-hist-O-ion}
\end{figure}

\begin{figure}
\centering
        \resizebox{\hsize}{!}{\includegraphics{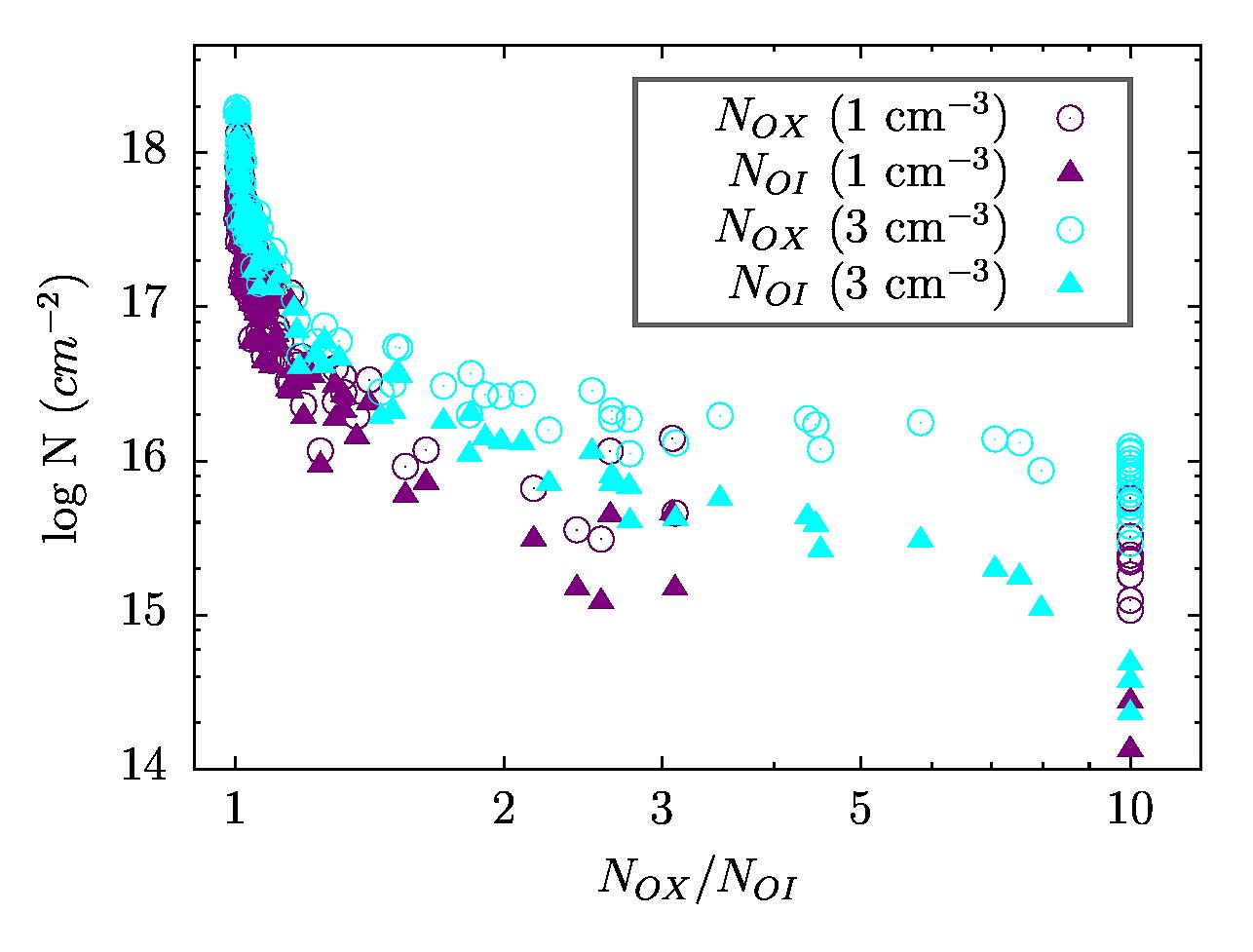}}
        \caption{$N_{OX}$ (solid points) and $N_{OI}$ (open points) as a function of absorption excess for 100 sightlines crossed through our ISM data cubes with densities 1 cm$^{-3}$ (purple points) and 3 cm$^{-3}$ (blue points).}
        \label{fig6:NOX-NOI-frac}
\end{figure}

\begin{table*}
\caption{Column densities of highly ionised gas \ion{C}{iv}, \ion{N}{v}, \ion{O}{vi} measured in GRB afterglow spectra ~\citep{Fox2008AA, Prochaska2008ApJ} and along the lines of sight through our turbulent ISM data cubes(Avg., Max). Only the measured absorption from low velocity components (few 100 km/s) are given. The column densities for the ions from ~\citet{Fox2008AA} without error bars correspond to saturated lines and are uncertain.}
\label{Table:1} 
\centering 
\begin{tabular}{c c c c c c c} 
\hline\hline 
GRB & z & $N_{CIV}$ & $N_{NV}$ & $N_{OVI}$ & $N_{SiIV}$ & Ref.\\ 
\hline 
Avg. (1 cm$^{-3}$)  & ... & 13.40 & 12.82 & 13.68  & 12.65 & this work \\ 

Max. (1 cm$^{-3}$) & ... & 14.37 & 13.59 & 14.19  & 13.86 & this work \\

Avg. (3 cm$^{-3}$)  & ... & 13.45 & 13.03 & 13.91 & 12.70 & this work \\

Max. (3 cm$^{-3}$) & ... & 14.32 & 14.34 & 14.44 & 13.64 & this work \\

GRB 021004 & 2.32 & ...  & 14.53$\pm $0.06 & 14.95$\pm $0.20  & 15.25$\pm $0.20 & 1\\ 
          & & ... & 14.64$\pm $0.04 & ... & ... & 2 \\ 

GRB 050730 & 3.97 & $\sim$ 15.24 & 13.80$\pm$0.14 & $\sim$ 15.94 & 14.16$\pm $0.06 & 1\\ 
         & & ... & 14.09$\pm$0.08 & ... & ... & 2\\ 

GRB 050820 & 2.62 & 13.13$\pm $0.06 & 13.17$\pm $0.08  & ... & 14.91$\pm $0.07 & 1 \\ 
                &  & ... & 13.45$\pm $0.05 & ... & ... & 2 \\  

GRB 050922C & 2.2 & 15.69$\pm $0.35 & 13.76$\pm $0.07 & 14.44$\pm $0.05 &14.49$\pm $0.16 & 1\\ 
                & & ... & $>$ 14.19 & ... & ... & 2 \\ 
GRB 060206 & 4.05 & ...  & 13.73$\pm $0.15 & ... & ... & 2\\ 

GRB 060607 & 3.08 & 15.94$\pm $0.32 & ... & ... & 13.54$\pm $0.09 & 1\\ 
                & & ... & $<$ 12.61 & ... & ... & 2 \\ 

GRB 071031 & 2.69 & $\sim$ 15.11 & 14.41$\pm $0.15 & $\sim$ 15.14 & 15.93$\pm $0.51 & 1\\ 

GRB 080310 & 2.42 & $\sim$ 16.66 & 14.03$\pm $0.05 & 15.12$\pm $0.09 & $\sim$ 15.00 & 1\\ 
\hline 
\end{tabular} 
\tablebib{(1)~\citet{Fox2008AA}; (2) ~\citet{Prochaska2008ApJ}.} 
\end{table*} 

Figures~\ref{fig3:CNO-ion-1cc} and ~\ref{fig4:CNO-ion-3cc} show the distribution of C, N, and O ion column densities traced along our 100 lines of sight for datacube1 and datacube3, respectively. For both the simulations, the column densities of ultra-highly ionised species \ion{C}{v}, \ion{N}{vi} and \ion{O}{vii} are found to be similar to the neutral and singly ionised components within the \ion{H}{ii} regions (gas T $> 10^{4}$ K). However, the column densities of \ion{C}{i}, \ion{N}{i}, and \ion{O}{i} increase further once the neutral ions within denser regions (gas T $< 10^{4}$ K) are considered. The column densities of \ion{C}{v}, \ion{N}{vi}, and \ion{O}{vii} are generally seen to be the highest after \ion{C}{i}, \ion{N}{i}, \ion{O}{i} and \ion{C}{ii}, \ion{N}{ii}, and \ion{O}{ii}, respectively. The denser data cube shows a rise in the ionisation profile at ionisation levels higher than \ion{C}{v}, \ion{N}{vi}, and \ion{O}{vii} owing to the higher volume fraction of hotter $\ge 10^{7}$ K gas in the ISM.

In Fig.~\ref{fig5:O-hist-O-ion} we plot the cumulative distribution of the soft X-ray absorption excess, $\frac{N_{OX}}{N_{OI}}$, measured along our 100 sightlines for datacube1 (green line) and datacube3 (brown line), and for comparison we also show the cumulative distribution of the excess measured in a sample of 16 GRBs (blue small-dashed line) ~\citep{Schady2011AA, Hartoog2013MNRAS, Sparre2014ApJ, DElia2014AA, Kruehler2013AA, Friis2015MNRAS}. We find that 92$\%$ and 72$\%$ of the sightlines through datacube1 and datacube3, respectively, trace $\frac{N_{OX}}{N_{OI}} < 4$, and $8 \%$ and $21 \%$ of the sightlines trace $\frac{N_{OX}}{N_{OI}} > 10$. The observed $\frac{N_{OX}}{N_{OI}}$ distribution is made up of the sample of GRBs with a detected X-ray afterglow and a well-measured \ion{Zn}{ii} column density from UV/optical spectra, which are typically undepleted elements. Assuming solar relative abundances~\citep{Asplund2009ARAA}, we converted the X-ray absorbing column density to an equivalent Zn column density following the method of ~\citet{Schady2011AA}, and we assume that $\frac{N_{ZnX}}{N_{ZnII}}$ $\sim$ $\frac{N_{OX}}{N_{OI}}$. We do not consider GRBs with reported column densities from other singly ionised metals, such as \ion{Si}{ii}, \ion{S}{ii} or \ion{Fe}{ii}, due to the uncertainty in the dust depletion ~\citep[e.g.][]{Savaglio2003ApJ, DeCia2013AA}. In Fig.~\ref{fig5:O-hist-O-ion}, it can be seen that 25$\%$ of the GRBs in our sample have a soft X-ray absorption excess $>10.0$ and 20$\%$ have an absorption excess $<4.0$.

All GRBs within our sample have a soft X-ray absorption excess $>2.0$ and thus warrant an additional absorbing component in addition to the neutral ISM. For the 25$\%$ of the GRBs with high excess $>10.0$, the turbulent ISM with density 3 cm$^{-3}$ could provide a natural explanation but the neutral and total oxygen column densities for such sightlines is 2-3 orders magnitude lower than those observed in GRB afterglow spectra ~\citep{Schady2011AA}. A high percentage of sightlines through our data cube show excess $< 2.0$, 86$\%$ and 63$\%$ for datacube1 and datacube3, respectively; these require an additional component to a turbulent ISM to produce the higher X-ray absorption excess. We find that the fraction of sightlines with significant absorption excess ($\frac{N_{OX}}{N_{OI}} > 2.0$) increases with density and the supernova rate, thus a turbulent ISM with higher density than 3 cm$^{-3}$ is a better representative of GRB host ISM.

In Fig.~\ref{fig6:NOX-NOI-frac} we show $N_{OX}$ and $N_{OI}$ against $\frac{N_{OX}}{N_{OI}}$ for the 100 sightlines crossed through the tested ISM data cubes. We see that for both the data cubes $N_{OX}$ remains fairly constant at $10^{15.5}$ cm$^{-2}$ for $2 \leq \frac{N_{OX}}{N_{OI}} \leq 10$, whereas $N_{OI}$ gradually decreases from $10^{16}$ cm$^{-2}$ at $\frac{N_{OX}}{N_{OI}} = 2$, to $10^{15}$ cm$^{-2}$ at $\frac{N_{OX}}{N_{OI}} = 7$ and shows a steep decline for $\frac{N_{OX}}{N_{OI}} \geq 8$. For datacube3 a high percentage $\sim 20\%$ show an absorption excess of $\frac{N_{OX}}{N_{OI}} \geq 10$, with $10^{15}$ cm$^{-2} \leq N_{OX} \leq 10^{16}$ cm$^{-2}$ and $N_{OI} < 10^{15}$ cm$^{-2}$. This suggests that sightlines with $\frac{N_{OX}}{N_{OI}} > 2$ cross a relatively smaller volume of neutral gas rather than a higher volume of ionised gas. The contribution from cold gas to both $N_{OX}$ and $N_{OI}$ dominates the measured $\frac{N_{OX}}{N_{OI}}$ ratio, such that any significant X-ray excess resulting from a sightlines through a hot, diffuse cloud is rapidly diluted away by the contribution from a cold, dense cloud. This is also reflected by the increase of 2-3 orders of magnitude in both $N_{OX}$ and $N_{OI}$ for sightlines with $\frac{N_{OX}}{N_{OI}} < 2$. 

In Table~\ref{Table:1} we compare \ion{C}{iv}, \ion{N}{v}, \ion{O}{vi}, and \ion{Si}{iv} column densities within our data cubes with those measured within a small sample of GRB host galaxies by~\citet{Prochaska2008ApJ} and~\citet{Fox2008AA}. To accomplish this, we track the Si ions in addition to the C, N, and O ions already traced. Si ions with two or more missing electrons are predominantly collisionally ionised, and thus in these cases our derived column densities are reliable. The ionisation potential of \ion{Si}{i} is lower than \ion{H}{i}, and thus we cannot accurately track the \ion{Si}{i} and \ion{Si}{ii} ion fractions and column densities within the data cubes. ~\citet{Fox2008AA} reported both high velocity (few 1000 km/s) and low velocity (few 100 km/s) absorption components, and in Table~\ref{Table:1} we report only the column densities of the low velocity absorption components, which are more likely to originate from the GRB host ISM. The velocity of the gas within the cells of our data cube is typically 10 - 100 km/s and is always below 500 km/s. We find that the measured column densities of \ion{C}{iv}, \ion{N}{v}, \ion{O}{vi,} and \ion{Si}{iv} are typically an order of magnitude larger than those traced along our 100 sightlines. For a typical GRB line of sight that crosses through a path length of 350 pc, the density of the GRB host ISM required to explain the column densities of \ion{C}{iv}, \ion{N}{v}, \ion{O}{vi,} and \ion{Si}{iv} ions should be an order of magnitude higher than the density of the ISM data cubes. Or else, the column densities can be produced by GRB host ISM with a density that is a factor of 2-3 higher than the ISM data cube and a path length of $\sim$ kpc. The later is the likely scenario given the small size of the GRB hosts ($\sim$ 5 kpc) ~\citep{Fruchter2006Nature}.

We report the $\frac{N_{CIV}}{N_{OVI}}$, $\frac{N_{NV}}{N_{OVI}}$, and $\frac{N_{CIV}}{N_{SiIV}}$ ion fractions from the data cubes, and GRB afterglow spectra for the low velocity absorption components in Table~\ref{Table:2}. We also report the median of the ion ratios along with the average ratios for our 100 sightlines. The average and median ion ratios of $\frac{N_{CIV}}{N_{OVI}}$ and $\frac{N_{NV}}{N_{OVI}}$ measured along sightlines through both the ISM data cubes are comparable to the observed ratios reported in ~\citet{Fox2008AA}. However, the average ion ratio of $\frac{N_{CIV}}{N_{SiIV}}$ is inconsistent with those observed in GRB afterglow spectra. Nevertheless, the median ratio of $\frac{N_{CIV}}{N_{SiIV}}$ for our sightlines through datacube1 and datacube3 are in agreement with $\frac{N_{CIV}}{N_{SiIV}}$ ion fraction observed in GRB afterglow spectra.

\section{Discussion}
The turbulent ISM with density 3 cm$^{-3}$ have properties that match the observed properties such as \ion{C}{iv}, \ion{N}{v}, \ion{O}{vi}, and \ion{Si}{iv} column densities, and $\frac{N_{CIV}}{N_{OVI}}$, $\frac{N_{NV}}{N_{OVI}}$, and $\frac{N_{CIV}}{N_{SiIV}}$ ion fractions. Sightlines passed through the ISM data cube with density 3 cm$^{-3}$ can produce the large soft X-ray absorption excess ($\frac{N_{OX}}{N_{OI}} > 10$) seen in GRB afterglow spectra. However, the total oxygen column densities for such sightlines is typically 10$^{16}$ cm$^{-2}$ and 2-3 orders of magnitude lower in comparison to those observed in GRB afterglow. While an ISM with higher density and larger length scale would increase $N_{OX}$ it is unlikely that these will produce the column densities observed in GRB spectra. Also, a large fraction of sightlines do not show significant absorption excess and this is in stark contrast to the observations. In the following section we explore additional consequences such an absorbing medium has on the observations profiles that either enhance or challenge the proposed scenarios and discuss alternative sources of X-ray absorption that could be responsible for the excess X-ray absorption.

\begin{table*}
\caption{Column density ratios for low velocity component of the highly ionised gas observed in the optical spectra of four GRBs taken from ~\citet{Fox2008AA}, and the average ratios for the 100 sightlines crossed through our data cubes. The values in bracket for $N_{CIV}/N_{OVI}$, $N_{NV}/N_{OVI}$, and $N_{CIV}/N_{SiIV}$ are the median ion ratios of our sightlines through the data cubes.}
\label{Table:2}
\centering
\begin{tabular}{c c c c c}
\hline\hline
GRB & z & $\frac{N_{CIV}}{N_{OVI}}$ & $\frac{N_{NV}}{N_{OVI}}$  & $\frac{N_{CIV}}{N_{SiIV}}$ \\ \\
\hline
GRB 050730 & 3.97  & 0.40 $\pm $ 0.25 & $<$0.09 & 3.7 $\pm $ 0.9 \\
GRB 050922C & 2.2 & ... & 0.21 $\pm $ 0.04 & ...\\
GRB 071031 & 2.69  & 1.6 $\pm$ 0.4 & $<$ 0.11  & 3.2 $\pm $ 0.9 \\
GRB 080310 & 2.42  & 0.55 $\pm $ 0.11 & $<$ 0.05 & 25 $\pm $ 4 \\
ISM (1 cm$^{-3}$) & -- & 0.63 (0.38) & 0.15 (0.125) & 60 (7.72)\\
ISM (3 cm$^{-3}$) & -- & 0.55 (0.25) & 0.16 (0.09) &1650 (7.75)\\
\hline
\end{tabular}
\end{table*}

\subsection{Could molecular hydrogen account for some of the excess?}
\begin{figure}
\centering
        \resizebox{\hsize}{!}{\includegraphics{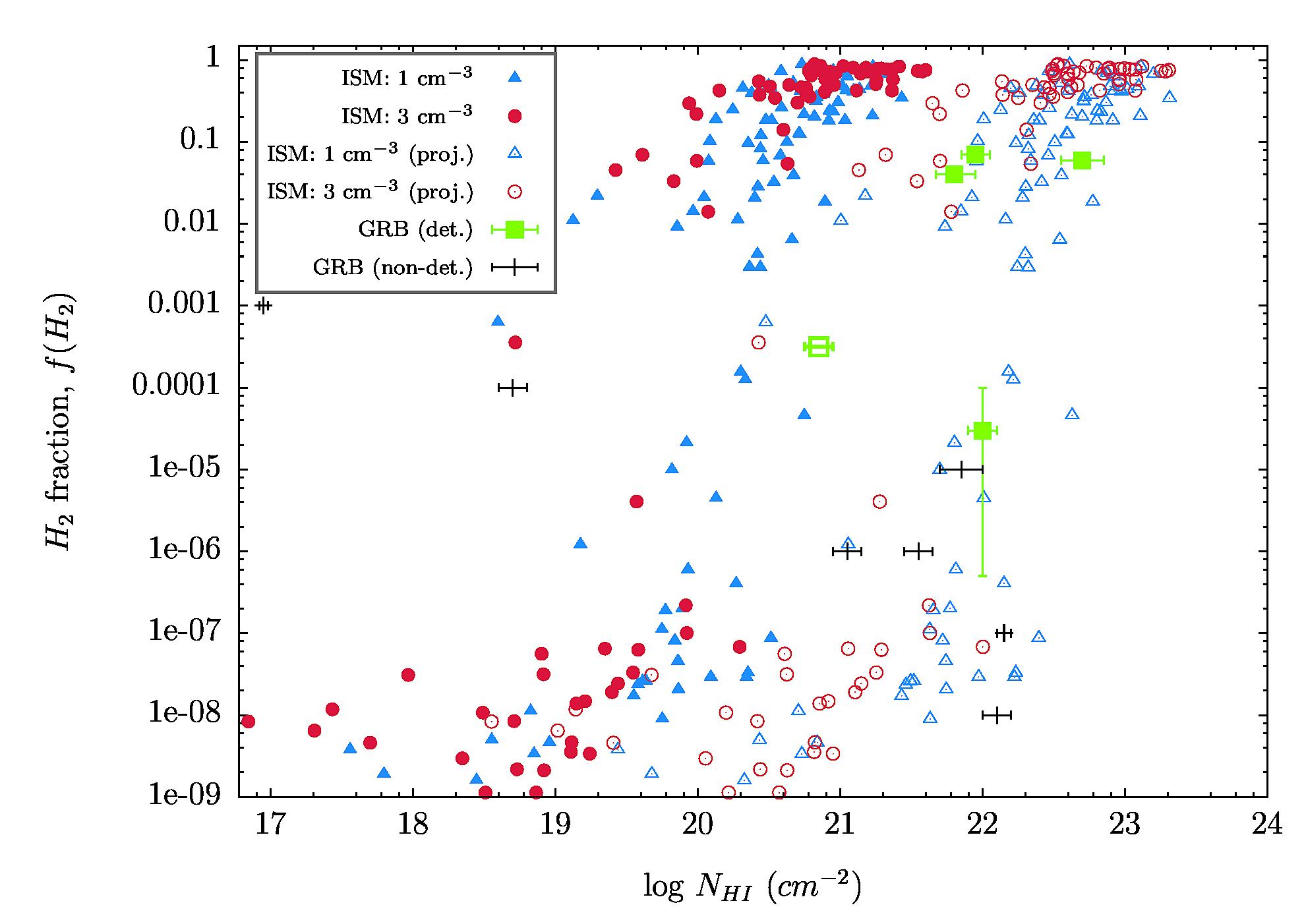}}
        \caption{Molecular hydrogen fraction ($f(H_{2})$) against $N_{HI}$ along the lines of sight for data cubes with densities 1 cm$^{-3}$ (blue solid triangles) and 3 cm$^{-3}$ (red solid circles). Column density of $H_{2}$, detected in GRB afterglow spectra for five GRBs (GRB 060206; ~\citep{Fynbo2006AA}, GRB 080607; ~\citet{Prochaska2009ApJ}, GRB 120327A; ~\citet{DElia2014AA}, GRB 120815A; ~\citet{Kruehler2013AA}, GRB 121014A; ~\citet{Friis2015MNRAS}) are indicated in green and six GRBs with upper limits on $f(H_{2})$ are indicated in black ~\citep{Tumlinson2007ApJ, Ledoux2009AA}. Projected $N_{HI}$ for the $f(H_{2})$ for data cubes with densities 1 cm$^{-3}$ (blue open triangles) and 3 cm$^{-3}$ (red open circles).}
        \label{fig8:MolH}
\end{figure}

Since GRB host galaxies are star-forming galaxies, and given the positive relation between molecular hydrogen and star formation ~\citep[i.e. Kennicutt-Schmidt law;][]{Kennicutt1998ApJ, Bigiel2008AJ}, we would expect there to be an abundance of molecular gas within GRB host galaxies. We therefore consider whether molecular hydrogen could cause a notable amount of additional X-ray absorption.

Our ISM data cubes include molecular gas and thus allow us to estimate the molecular hydrogen fraction, $f(H_{2}) = 2N_{H_{2}}/\left[2N_{H_{2}} + N_{HI}\right]$, along the traced line of sights. In Fig.~\ref{fig8:MolH} we show $f(H_{2})$ against $N_{HI}$ for the 100 sightlines crossed through our ISM data cubes indicated by blue solid triangles and red solid circles for datacube1 and datacube3, respectively, for five GRBs\footnote{The detection of $H_{2}$ along the line of sight to GRB 060206 is tenuous ~\citep{Prochaska2009ApJ}.} (green squares) with detected $H_{2}$ absorption in their afterglow spectra and six GRBs (black points) with upper limits on column densities of $H_{2}$ within the host ISM. As expected, for both ISM data cubes, we find that the sightlines with larger neutral metal column densities encounter a higher $H_{2}$ fraction. The observed \ion{H}{I} column densities are larger than those measured along our sightlines through both data cubes. As noted earlier, this is due to the larger length scales probed by the GRB afterglow and/or higher density of the GRB host ISM. We scale the path of the typical path length of our sightlines to increase \ion{H}{I} column densities for the sightlines within our data cubes to form an overlap between the detections from the GRB afterglow spectra and those are indicated with blue open triangles and red open circles for datacube1 and datacube3, respectively. For each data cube, the projected \ion{H}{I} column densities are obtained by multiplying by a scale factor, which is given by the ratio of average $N_{HI}$ for three GRBs with $0.01 < f(H_{2}) < 0.1 $ to the $N_{HI}$ within our sightlines at average $f(H_{2})$ for the three GRBs. The scaled path lengths for datacube1 is $\sim$26 kpc and for datacube3 is $\sim$17 kpc, assuming typical path length of 350 pc for our sightlines. 

Molecular hydrogen can absorb soft X-rays and the effective absorption excess can be estimated as $[2.85 N_{H_{2}} + N_{HI}]/N_{HI}$, where the cross-section of $H_{2}$ is 2.85$\sigma_{H}$ ~\citep{Wilms2000ApJ}. In order for $H_{2}$ to produce a soft X-ray absorption excess of $>$ 2.0 , the $H_{2}$ fraction would then need to be $>$ 0.5 . For all GRBs with detected $H_{2}$ absorption, the molecular hydrogen fraction is less than $0.1$, and this corresponds to a soft X-ray absorption excess of less than 1.2 . Furthermore, for a sample of six GRBs with good quality optical afterglow spectra, ~\citet{Tumlinson2007ApJ} place an average upper limit of $10^{-6.5}$ on $f(H_{2})$. On the basis of simulations,  ~\citet{Whalen2008ApJ} show that the molecular hydrogen within the GRB vicinity would be destroyed by UV radiation from massive stars prior to the onset of GRB. They  highlight the possibility, however, that absorption from molecular clouds from less active star-forming regions within the GRB host that are unrelated to the GRB progenitor region may be detected occasionally in the GRB afterglow spectra. So far, such examples of intervening molecular clouds have not been the cause for the X-ray absorption excess. For example, GRB 120327A ~\citep{DElia2014AA}, GRB 120815A ~\citep{Kruehler2013AA}, and GRB 121014A ~\citep{Friis2015MNRAS} with firm detections of $f(H_{2})$ and three of the GRBs in ~\citet{Tumlinson2007ApJ} (GRB 050820,GRB 071031, and GRB 080413) that have stringent upper limits on $f(H_{2})$, are part of our sample and they have significant X-ray absorption excess ($N_{OX}/N_{OI} >$ 4.0).  

Although, our analysis suggests that a large $H_{2}$ fraction ($f_{H2} >$ 0.5) would result in an observed X-ray excess, the excessive attenuation from gas and dust along such sightlines would typically prevent the GRB afterglows from being detected. Such sightlines could be observed in large GRB hosts with a dense ISM and with early-time sensitive spectra. These shall also lead to detection of high hydrogen fractions ($f(H_{2}) > 0.1$), which have been missed in observations so far ~\citep{Noterdaeme2015AA}. 

Interstellar dust is unlikely to be the cause of the X-ray excess. The relation between $A_{V}$ and $N_{HX}$ along GRB lines of sight has been  studied extensively. The seminal works of ~\citet{GalamaWijers2001ApJ} and ~\citet{Stratta2004ApJ} were the first to point out the very high gas-to-dust ratio, $\frac{N_{HX}}{A_{V}}$, in comparison to the Milky Way. Studies with larger samples that include heavily dust-extinguished GRBs verify this claim and have provided evidence of a positive correlation between $N_{HX}$ and $A_{V}$ ~\citep{Schady2007MNRAS, Starling2007ApJ, Schady2010MNRAS, Greiner2011AA, Zafar2011AA, Zafar2012ApJ, Watson2013ApJ, Covino2013MNRAS}. The most heavily dust extinguished GRBs are found in the most massive GRB hosts with high $N_{HX}$ and $N_{HI}$ ~\citep{Kruehler2011AA, Perley2013ApJ}, and although $N_{HX}$ and $A_{V}$ appear to be correlated, the rate of increase in $N_{HX}$ does not trace the increase in $A_{V}$, with $\frac{N_{HX}}{A_{V}}$ decreasing with an increase in $A_{V}$ ~\citep{Kruehler2011AA}. GRB 080607, a roughly solar metallicity GRB, is a good example, having had one of the dustiest sightlines yet detected ~\citep[$A_{V} \sim$ 3.3;][]{Perley2011AJ}, and yet a similar $N_{HX}$ and $N_{HI}$. Furthermore, most of the GRBs in the sample of ~\citet{Schady2011AA} were relatively unextinguished by dust. Thus, dust within the GRB host ISM cannot be the source of the soft X-ray absorption excess.

\subsection{Additional sources of soft X-ray absorption excess in GRBs}
As already stated in section 1, the alternate sources of X-ray absorption, such as as a dense circumburst medium ~\citep{Krongold2013ApJ}, or hot IGM ~\citep{Behar2011ApJ,Starling2013MNRAS} face challenges. ~\citet{Krongold2013ApJ} require the circumburst medium to be dense ($> 100$ cm$^{-3}$), pre-ionised and to extend to a few 10 pc, and how such dense hot regions of gas could exist in pressure equilibrium with the neutral phase ISM is not clear. From the density vs temperature plots for multiphase ISM, left panel in Fig.~\ref{fig1:ISM-1cc} and Fig.~\ref{fig2:ISM-3cc}, we do not find gas clouds with density $>$ 10 cm$^{-3}$ and temperature $> 10^{4}$ K in a multiphase ISM. Similar conditions might however be present in young massive star clusters ~\citep[e.g.][]{Krause2012AA, Palous2013ApJ} and these scenarios need to be pursued further. 

Highly ionised gas, unaccounted for in our simulations, could also arise as a result of photoionisation from massive stars and delayed recombination. Photoionisation by hot stars is unsuitable to produce the high ionisation stages of oxygen (\ion{O}{vi} and \ion{O}{vii}),which we are primarily interested in. Soft X-ray emission (0.2 - 2 keV) from SN-remnants, superbubbles, and non-thermal sources ~\citep[e.g.][]{Sasaki2002AA, Haberl2012AA, Krause2014AA} scales with the star formation rate and would total to a luminosity of 10$^{40}$ erg s$^{-1}$ for a typical GRB host, insufficient to produce significant columns of hot gas. Non-equilibrium ionisation (NEI) may change the detailed ionisation structure compared to the collisional ionisation assumption made here ~\citep{Avillez2012ApJ}. NEI can cause ions like \ion{C}{v}, \ion{N}{vi}, and \ion{O}{vi} to be produced between $10^{3.8} \le$ T $\le 10^{5.5}$ K unlike CIE conditions, where these ions can only be produced at T $> 10^{5}$ K. However, the soft X-ray absorption excess is measured as the ratio of total oxygen column density to the neutral oxygen column density and, thus, is insensitive to the detailed ionisation structure. We find neutral oxygen in denser regions for a collisionally ionised ISM, where collisions are frequent and delayed recombination does not have a large impact. Therefore, while the detailed NEI stage abundances would be interesting to have in this context, they would not lead to very different values for NOX and NOI. Therefore, we believe the conclusions of our analysis are robust. 

Some source of hot gas, in addition to the turbulent ISM, is required to explain the observed X-ray absorption excess, and in particular, the paucity of GRBs in our sample with low soft X-ray absorption excess, $\frac{N_{OX}}{N_{OI}} < 2.0$. Absorption by collisionally ionised ISM eases the constraints on absorption by the circumburst medium and the WHIM. 

\section{Summary and future work}
We estimate the soft X-ray absorption excess for a sample of 16 GRBs as the fraction of total oxygen column density to neutral oxygen column density, $\frac{N_{OX}}{N_{OI}}$, and we find that significant soft X-ray absorption excesses, $\frac{N_{OX}}{N_{OI}} > 2.0$, occurs in every case. Furthermore, half of the GRBs in our sample show $\frac{N_{OX}}{N_{OI}} > 10$. It is important to quantify the soft X-ray absorption for GRBs to estimate and explain the total X-ray absorbing column densities that produce this excess. In our simulations of collisionally ionised ISM, typical of star-forming galaxies such as GRB hosts, we find values of excess compatible with observations only for a small number of sightlines. In these cases, however, $N_{OX}$ is orders of magnitude below the observed values, which are typically of the order of 10$^{18.5}$ cm$^{-2}$. We therefore find that the ISM in GRB host galaxies cannot be the main contributor to the X-ray absorbing column for GRB afterglows. Alternate scenarios, possibly involving absorption from the circumburst medium and the WHIM, will have to be invoked to explain soft X-ray absorption excess along all the sightlines.

To obtain the observed \ion{H}{i} column densities, the path length of the sightlines through datacube1 is required to be $\sim$26 kpc and the path length of the sightlines through datacube3 is required to be $\sim$17 kpc. GRB hosts typically are smaller in size, $<$10 kpc ~\citep{Fruchter2006Nature}, and thus their ISM are required to be denser by a factor of 2 or 3 over 3 cm$^{-3}$ to explain the observed \ion{H}{i} column densities. ISM data cubes with density 10 cm$^{-3}$  have a higher SN rate (77 Myr$^{-1}$) and thus a more turbulent ISM. We observe that increase in density of the ISM data cubes results in higher volume fraction of hot gas and, thus, an ISM with density 10 cm$^{-3}$ could have a lower fraction of sightlines with $\frac{N_{OX}}{N_{OI}} < 2.0$. It seems, however, unlikely, extrapolating from the two simulations we have analysed, that the total oxygen column would increase enough to match observations.

On the basis of the ISM data cubes in our analysis, we predict the presence of ultra-highly ionised gas traced by \ion{C}{V}, \ion{N}{VI,} and, \ion{O}{VII}. In Chandra and XMM-Newton X-ray observations of the Blazar outburst, K-$\alpha$ absorption lines from elements such as \ion{N}{VI} and \ion{O}{VII} produced by the WHIM have been detected, corresponding to column densities of $\sim$$10^{15} - 10^{16}$ cm$^{-2}$, albeit with low statistical significance ~\citep[e.g.][]{Nicastro2005ApJ, Kaastra2006apJ}. The GRB afterglow soft X-ray, $0.3 -10$ keV, integrated flux at beginning of the Swift XRT observations is $\sim$$10^{-10}$ erg/cm$^{2}$/s. This flux typically decays to $\sim$$10^{-12}$ erg/cm$^{2}$/s after about six hours and to $\sim$$10^{-13}$ erg/cm$^{2}$/s about 24 hours after the GRB trigger. The next generation of X-ray observatories such as Astro-H ~\citep{Takahashi2012SPIE} and the Athena X-ray Observatory ~\citep{Nandra2013arXiv} will be able to detect absorption features from ultra-highly ionised gas within the WHIM to high significance even approximately 10 hours after the GRB trigger. Such a detection with GRBs serving as beacons would be possible in $\sim$$100$ ks with Astro-H ~\citep{Tashiro2014arXiv} and in $\sim$$50$ ks with X-ray IFU ~\citep{Nicastro2014cosp} on board Athena ~\citep{Barret2013arXiv}. The regions of hot gas within the GRB host ISM are expected to be denser and localised within sub-kpc scales in comparison to the diffuse WHIM; the absorption signature due to the ultra-highly ionised gas within the ISM and circumburst medium should be easily detected within the exposure times required to detect the WHIM. Thus, apart from probes of the WHIM, GRBs shall serve as unique probes of the hot gas in star-forming galaxies that are ionised by massive stars.

\begin{acknowledgements}
We thank D. Breitschwerdt, D. B. Fox, and H. van Eerten for their suggestions and comments. P.S. and M.T. acknowledge support through the Sofja Kovalevskaja Award to P. Schady from the Alexander von Humboldt Foundation of Germany. This work was supported by funding from Deutsche Forschungsgemeinschaft under DFG project number PR 569/10-1 in the context of the Priority Program 1573 “Physics of the Interstellar Medium”.

\end{acknowledgements}

\bibliography{Reference}

\end{document}